\pdfoutput=1
\documentclass[a4paper,fleqn,usenatbib]{mnras}

\usepackage{newtxtext,newtxmath}

\usepackage[T1]{fontenc}
\usepackage{ae,aecompl}

\usepackage{graphicx}
\usepackage{amssymb}

\usepackage{cases}
\usepackage{hyperref}
\usepackage{wasysym}
\usepackage{aas_macros}

\newcommand{\rcb}{\mathrm{rcb}}

\newcommand{\htwoo}{\mathrm{H_2 O}}
\newcommand{\hhe}{\mathrm{H/He}}
\newcommand{\cotwo}{\mathrm{CO_2}}

\title[Effects of composition on impact-driven atmospheric loss]{Losing Oceans: The Effects of Composition on the Thermal Component of Impact-driven Atmospheric Loss}
\author[J. B. Biersteker and H. E. Schlichting]{John B. Biersteker$^{1}$\thanks{Email: jo22395@mit.edu} and 
Hilke E. Schlichting$^{1,2}$
\\
$^{1}$Massachusetts Institute of Technology, 77 Massachusetts Avenue, Cambridge, MA 02139-4307, USA\\
$^{2}$UCLA, 595 Charles E. Young Drive East, Los Angeles, CA 90095, USA
}

\date{Accepted XXX. Received YYY; in original form ZZZ}

\pubyear{2020}

\begin{document}
\label{firstpage}
\pagerange{\pageref{firstpage}--\pageref{lastpage}}
\maketitle

\begin{abstract}
The formation of the solar system's terrestrial planets concluded with a period of giant impacts. Previous works examining the volatile loss caused by the impact shock in the moon-forming impact find atmospheric losses of at most 20--30 per cent and essentially no loss of oceans. However, giant impacts also result in thermal heating, which can lead to significant atmospheric escape via a Parker-type wind. Here we show that $\htwoo$ and other high-mean molecular weight outgassed species can be efficiently lost through this thermal wind if present in a hydrogen-dominated atmosphere, substantially altering the final volatile inventory of terrestrial planets. Specifically, we demonstrate that a giant impact of a Mars-sized embryo with a proto-Earth can remove several Earth oceans' worth of $\htwoo$, and other heavier volatile species, together with a primordial hydrogen-dominated atmosphere. These results may offer an explanation for the observed depletion in Earth's light noble gas budget and for its depleted xenon inventory, which suggest that Earth underwent significant atmospheric loss by the end of its accretion. Because planetary embryos are massive enough to accrete primordial hydrogen envelopes and because giant impacts are stochastic and occur concurrently with other early atmospheric evolutionary processes, our results suggest a wide diversity in terrestrial planet volatile budgets.
\end{abstract}

\section{Introduction}
The growth of rocky planets in the solar system is believed to have culminated in a series of giant impacts where planetary embryos, bodies which are thousands of kilometers in size, collide and merge over a timescale of ${\sim}10\text{--}100$ million years to form the terrestrial planets \citep{2012AREPS..40..251M}. This giant impact phase is a natural outcome of the growth of planetary embryos in the stabilizing presence of a gas disk. While the gas is present, interactions between the embryos and the gas damp the embryos’ orbital eccentricities. When the gas is removed, gravitational interactions between the planetary embryos lead to excitation of their orbital eccentricities, resulting in orbit crossing and collisions \citep[e.g.,][]{1998Icar..136..304C, 2019Icar..329...88W}. This period of instability has an estimated duration of ${\sim}30{-}100~\mathrm{Myr}$ at ${\sim}1~\mathrm{au}$ \citep{2012AREPS..40..251M, 2015MNRAS.453.3619I, 2019Icar..329...88W}. In the case of Earth, bulk accretion was likely completed with the moon-forming impact ${\lesssim}60~\mathrm{Myr}$ after the formation of the solar system \citep{2001Natur.412..708C, 2014AREPS..42..551A, 2017SciA....3E2365B}.

The atmospheres of young terrestrial planets evolve significantly over their first $100\text{--}200 ~\mathrm{Myr}$. Isotopic evidence indicates that the formation timescale for Mars is comparable to the lifetime of the gas disk \citep{2011Natur.473..489D}, suggesting that embryos that formed the terrestrial planets grew to at least the mass of Mars (${\sim}0.1~M_\oplus$) prior to the dissipation of nebular gas. 
Mars-sized and larger embryos are massive enough to directly accrete a primary atmosphere of hydrogen and helium ($\hhe$) from the disk \citep{1979E&PSL..43...22H, 2014MNRAS.439.3225L, 2015MNRAS.448.1751I}.
The noble gas composition in the Earth's mantle may reflect the dissolution of nebular gas from such a primordial atmosphere into the proto-Earth's magma ocean \citep{1996Sci...273.1814H, 2012Natur.486..101M, 2019Natur.565...78W}. In addition, the ubiquity of sub-Neptunes\textendash exoplanets residing within $1~\mathrm{au}$ of their host stars that are shrouded in hydrogen-dominated envelopes containing roughly 1 per cent of the planet's total mass\textendash suggests that rapid core formation and gas accretion from the primordial disk is commonplace in planet formation \mbox{\citep{2013ApJ...766...81F,2016ApJ...825...29G}}. 
In some planet formation models \citep{2014Icar..239...74O, 2017AREPS..45..359J}, the embryos that eventually form the terrestrial planets can grow to ${\gtrsim}50$ per cent of their final size before the dissipation of the gas disk. It is therefore possible that planetary embryos could have not only accreted but also retained significant $\hhe$ envelopes over $10\text{--}100~\mathrm{Myr}$ timescales, well into the giant impact phase \citep{2014MNRAS.439.3225L}.

In addition to directly accreting an atmosphere, as a planet cools and its magma ocean solidifies, volatiles that were dissolved in the magma ocean can be outgassed. The magma oceans of terrestrial planets without substantial primary atmospheres can cool quickly ($1\text{--}10~\mathrm{Myr}$), likely producing an episode of catastrophic outgassing near the end of the magma ocean's crystallization that releases a secondary atmosphere dominated by water vapor ($\htwoo$) and carbon dioxide ($\cotwo$) \citep{2008E&PSL.271..181E,2012AREPS..40..113E}. This steam atmosphere can persist for ${\sim}10~\mathrm{Myr}$ before cooling and condensing \citep{2008E&PSL.271..181E, 2011Ap&SS.332..359E}. If instead the magma ocean is blanketed by an optically thick hydrogen-dominated envelope, the added insulation will keep surface temperatures high, allowing a magma ocean to persist through the giant impact phase. Chemical equilibration between the long-lived magma ocean and the atmosphere can produce mixed atmospheres of predominantly $\hhe$ and $\htwoo$, with the relative abundances determined by the redox state of the magma ocean \citep{2020ApJ...891..111K}.

The young atmospheres of terrestrial planets are subjected to loss processes beyond those associated with impacts. In particular, stellar radiation-powered hydrodynamic escape of atmospheric hydrogen, either nebular in origin or produced through photodissociation of $\htwoo$, is thought to partially explain the observed variation in volatile abundance and noble gas mass fractionation in the solar system's terrestrial planets \citep{1986Icar...68..462Z,2018Icar..307..327O, 2018A&ARv..26....2L}. Depending on the assumed high energy flux from the host star, primary and secondary atmospheres can be significantly eroded or even lost over $10\text{--}100~\mathrm{Myr}$ timescales \citep{2015ApJ...815L..12J, 2014MNRAS.439.3225L, 2018Icar..307..327O}. 
Given the timing and duration of these key processes driving atmospheric evolution, the giant impact phase likely spans a wide diversity of atmospheric conditions on young terrestrial planets; an early impact may occur on a planet with a thick $\hhe$ envelope, while an impactor arriving near the end of the giant impact phase may encounter a much less massive secondary atmosphere.

Giant impacts have the potential to dramatically alter the atmosphere of an accreting planet. Past work examining volatile loss in giant impacts focused on the loss triggered directly by the impact shock \citep{2003Icar..164..149G, 2005Natur.433..842G, 2015Icar..247...81S}. Here, we determine the atmospheric loss as a result of the thermal heating caused by the giant impact, which can drive atmospheric escape via a Parker-type wind \citep{1997Icar..126..148P}. 
It has already been demonstrated that this thermal loss can lead to complete depletion of primordial $\hhe$ envelopes during the formation of super-Earths and sub-Neptunes and that, for such planets, this thermal loss component exceeds that due to the impact shock by an order of magnitude \citep{2019MNRAS.485.4454B}.  Here, we examine the effect of atmospheric composition on this impact-triggered thermal loss, with a focus on mixed atmospheres, consisting of primordial and outgassed material, and on secondary atmospheres, which are composed of heavier volatile species. In Section \ref{sec: atmosphere model} we review the model of atmospheric evolution and loss following an impact and present resulting analytical estimates for conditions where substantial atmospheric loss can be expected. In Section \ref{sec: terran numerical results}, we present numerical simulations for four case studies spanning a range from pure $\hhe$ envelopes to completely secondary steam atmospheres. We close in Section \ref{sec: terran discussion} with a discussion.

\section{Atmospheric evolution model}
\label{sec: atmosphere model}
We calculate the extent of atmospheric loss from planetary winds following an impact using the model developed in \citet{2019MNRAS.485.4454B}, adapted to handle the diverse atmospheric compositions and the range of impact scenarios expected in terrestrial planet formation.
This model assumes a two-layer atmosphere: the lower layer is optically thick and convecting, as expected for envelopes with the masses and compositions considered here, and the upper layer is radiative and approximately isothermal \citep{2006ApJ...648..666R,2014ApJ...786...21P,2016ApJ...825...29G}. The boundary between these atmospheric layers is defined as the radiative-convective boundary (RCB) where $r = R_\rcb$ denotes the radial distance from centre of the planet to the radiative-convective boundary. Because these planets are young and possess insulating envelopes, we assume a molten surface that efficiently exchanges heat with the atmosphere. Atmospheric mass loss occurs through hydrodynamic outflow beyond the outer radius ($R_\mathrm{out}$), which is the smaller of the Hill radius ($R_H = a (M_p / 3 M_\mathrm{star})^{1/3}$) and the Bondi radius ($R_B = 2 G M_p / c_s^2$), where $M_p$ is the planet mass, $M_\mathrm{star}$ is the mass of the host star, $a$ is the planet's orbital radius, $c_s$ is the isothermal sound speed of the gas, and $G$ is the gravitational constant. The mass loss rate is limited by either the density of gas at the outer radius, or the rate at which gas can be delivered from the convective region to the outer radius. 
Following a giant impact, some of the energy deposited by the impactor heats the planet, causing the envelope to thermally expand, leading to accelerated atmospheric escape.

\subsection{Atmospheric structure}
We model the inner convective region of the envelope with an adiabatic profile defined by
\begin{align}
\frac{\rho}{\rho_p} & = \left[ \frac{\gamma - 1}{\gamma} \Lambda \left( \frac{R_p}{r} - 1 \right) + 1 \right]^{\frac{1}{\gamma - 1}} \text{,}
\label{eq: density}
\\
\frac{P}{P_p} & = \left[ \frac{\gamma - 1}{\gamma} \Lambda \left( \frac{R_p}{r} - 1 \right) + 1 \right]^{\frac{\gamma}{\gamma - 1}} \text{,}
\\
\frac{T}{T_p} & = \frac{\gamma - 1}{\gamma} \Lambda \left( \frac{R_p}{r} - 1 \right) + 1 \text{,}
\end{align}
where $R_p$ is the planet's radius, $\gamma$ is the adiabatic index of the atmosphere, and $\rho_p$, $P_p$, and $T_p$ are the density, pressure, and temperature at the surface of the planet. We define $\Lambda \equiv G M_p \mu m_p / (R_p k_B T_p)$, where $k_B$ is the Boltzmann constant, $m_p$ is the proton mass, and $\mu$ is the mean molecular weight of the atmosphere. 
Beyond the RCB, we model the atmosphere as isothermal, with a temperature given by the equilibrium temperature, $T_\mathrm{eq} \simeq  T_\mathrm{star} \sqrt{R_\mathrm{star} / 2 a} $, where $R_\mathrm{star}$ and $T_\mathrm{star}$ are the radius and effective temperature of the host star.
The density profile in this region is exponential,
\begin{align}
\label{eq: expo density}
\rho = \rho_\rcb \exp{\left[ \frac{R_\rcb}{h} \left( \frac{R_\rcb}{r} - 1 \right) \right]} \text{,}
\end{align}
where $h = k_B T_\mathrm{eq} R_\rcb^2 / G M_p \mu$ is the scale height. For the atmospheres considered here, the majority of the mass and energy is contained in the convective region of the envelope. These quantities can be obtained by integrating over the adiabatic profile:
\begin{align}
M_\mathrm{env} & = 4 \pi R_p^3 \rho_p \int_{R_p}^{R_\rcb} \rho r^2 dr
\\
E_g & = -4 \pi G M_p \rho_p R_p^2 \int_{R_p}^{R_\rcb} \rho r dr
\\
E_\mathrm{th} & = 4 \pi \rho_p R_p^3 \frac{k_B T_p}{\mu m_p (\gamma - 1)} \int_{R_p}^{R_\rcb} T r^2 dr \text{.}
\end{align}

Because we are interested in young (${\lesssim}100~\mathrm{Myr}$) planets that are insulated by optically thick envelopes, we make the simplifying assumption that the solid planet can be treated as fully molten and isothermal. We additionally assume that the molten surface efficiently exchanges heat with the overlying envelope so that the planet's temperature is roughly equal to the base temperature of the envelope, $T_p$. The thermal energy of the planet is then $E \sim c_{V,p} M_p T_p$ where we adopt $c_{V,p} \sim 7.5 \times 10^6 ~\mathrm{erg}~\mathrm{g}^{-1} \mathrm{K}^{-1}$ as the approximate heat capacity of the solid planet \citep{2001PhRvB..64d5123A, 1995ApJ...450..463G}. 

The atmospheric mass loss rate is determined by the atmospheric profile and is limited either by the density of gas at the outer radius or the rate at which gas can be resupplied to this radius from the RCB. The energy required to lift gas from $R_\rcb$ to $R_\mathrm{out}$ is provided by a combination of the cooling luminosity of the planet ($L_\rcb$) and incoming stellar radiation, yielding an additional limit on the mass loss rate. Combining these conditions yields
\begin{align}
\label{eq: mdot definition}
\left| \dot{M} \right| = \mathrm{min}{\left(4 \pi R_\mathrm{out}^2 \rho_\mathrm{out} c_s, \frac{L_\rcb R_\rcb}{G M_p} \right)}  \text{,}
\end{align}
where $\rho_\mathrm{out}$ is the density at $R_\mathrm{out}$, $c_s = \sqrt{\gamma k_B T / \mu m_p}$ is the sound speed, and the luminosity at the RCB is
\begin{align}
L_\rcb = \frac{\gamma - 1} {\gamma} \frac{64 \pi \sigma T_\rcb^3 G M_p \mu m_p}{3 \kappa_R \rho_\rcb k_B} \text{,}
\end{align}
where $\kappa_R$ is the Rosseland mean opacity. We adopt a Rosseland mean opacity of $\kappa_R = 0.1 ~\mathrm{cm}^2~\mathrm{g}^{-1}$ for pure $\hhe$ and mixed $\hhe$+$\htwoo$ atmospheres, and $\kappa_R = 0.05~\mathrm{cm}^2~\mathrm{g}^{-1}$ when considering $\htwoo$ atmospheres as appropriate approximations for the conditions at the RCB \citep{2008ApJS..174..504F, 2013ApJ...775...10V}. The energy lost through atmospheric mass loss is then $\dot{E}_{\mathrm{env}, m} \approx G M_p \dot{M} / R_\rcb$, while the radiative cooling is given by $\dot{E}_{\mathrm{env}, \mathrm{L}} = -L_\rcb$.

\begin{figure}
\centering
	\includegraphics[width=0.5\textwidth]{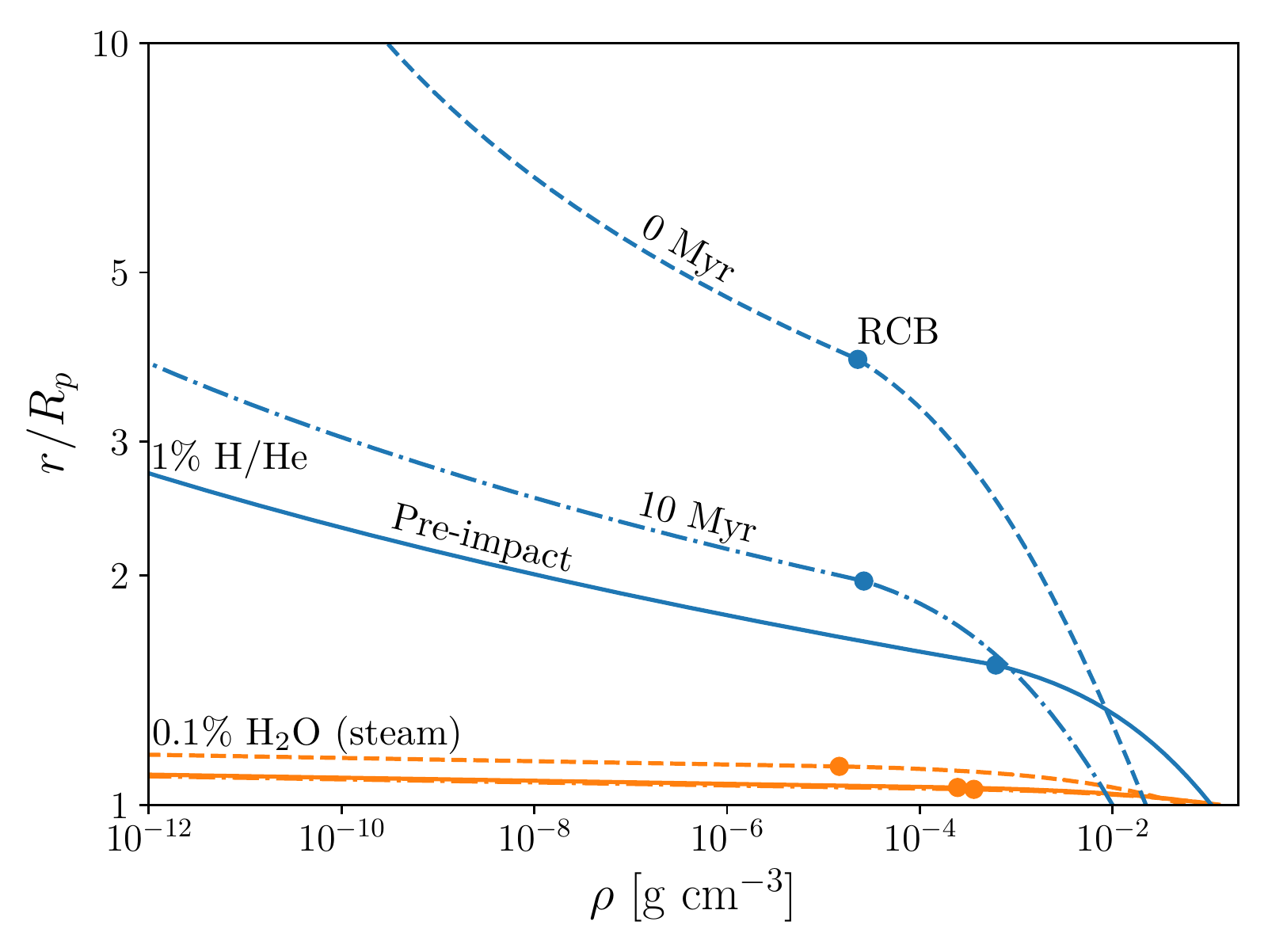}
	\caption{Examples of the evolution of atmospheric structure following an impact. Atmospheric density ($\rho$) at different times is shown as a function of radius ($r/R_p$) for a $\hhe$ atmosphere with a mass fraction of $1$ per cent (blue) and an $\htwoo$ (steam) atmosphere with a mass fraction of $0.1$ per cent (orange). Three times are shown: pre-impact (solid), soon after the impact and thermal re-equilibration of the planet (dashed), and $10 ~\mathrm{Myr}$ after impact (dash-dotted). Circles mark the radiative-convective boundary (RCB), where the density profile transitions from adiabatic to exponential. The thermal energy deposited by the impact causes the envelope to expand, increasing the density at high altitudes by orders of magnitude and promoting atmospheric loss. As mass is lost and the envelope cools, the envelope contracts and mass loss is quenched. The steam atmosphere, due to its higher mean molecular weight, experiences only minor inflation and no appreciable loss, while the hydrogen-dominated envelope expands dramatically and is reduced to ${\sim}10$ per cent of its original mass.}
	\label{fig: density profile evolution}
\end{figure}

We combine these models of the planet and its atmosphere to numerically calculate the thermal evolution and mass loss of the planetary atmosphere following an impact (see Figure \ref{fig: density profile evolution}). The total energy, $E = E_p + E_g + E_\mathrm{th}$ uniquely determines the atmospheric structure, which in turn defines the mass loss rate (Equation \eqref{eq: mdot definition}). The change in energy is then given by $\dot{E} = \dot{E}_{\mathrm{env}, m} + \dot{E}_{\mathrm{env}, \mathrm{L}}$. We numerically integrate these equations over $2~\mathrm{Gyr}$ using the \texttt{odeint} routine provided by \texttt{scipy} to determine the final atmospheric structure and cumulative atmospheric mass loss for a given planet. The model described above assumes that the atmosphere remains collisional at altitudes below $R_\mathrm{out}$. We check this condition is satisfied at each time step. We do this by comparing the altitude of the exobase $R_\mathrm{exo}$ to $R_\mathrm{out}$, where the exobase is conservatively defined as the location where $h$ (from Equation \eqref{eq: expo density}) is equal to the mean free path of a gas molecule. We find that halting atmospheric mass loss once the flow is no longer collisional has only a minor effect on the overall results. This is because the flow typically becomes collisionless when either the atmosphere has already been nearly completely lost, or when our model already predicts limited atmospheric mass loss.

The initial conditions for our calculations of atmospheric evolution are determined by the pre-impact state of the planet and the chosen impactor mass and velocity. We set an initial temperature for the base of the atmosphere of $T_{p, 0} = 2000~\mathrm{K}$, consistent with our assumption of a molten planetary surface and expectations for young planets hosting substantial insulating atmospheres. We assume the impact results in perfect accretion, so that $M'_p = M_p + M_\mathrm{imp}$, where the prime indicates the post-impact value. The energy added by the impact is $E_\mathrm{imp} \simeq \eta M_\mathrm{imp} v_\mathrm{imp}^2 / 2$, where $M_\mathrm{imp}$ and $v_\mathrm{imp}$ are the impactor mass and velocity, and $\eta$ describes the fraction of the impact energy that is used to heat the envelope and the core (where the core refers to the iron and silicate portion of the planet). In our calculations we assume $\eta \sim 0.5$ and $v_\mathrm{imp} \simeq \sqrt{2} v_\mathrm{esc}$, where $v_\mathrm{esc}$ is the mutual escape velocity. These values fall within the range of expectations for giant impacts during terrestrial planet formation \citep{2006Icar..184...39O, 2014Icar..239...74O, 2020JGRE..12506042C}.

\subsection{Maximum molecular weight for atmospheric loss}
\label{sec: max mu}
All of the planets considered here have molten cores with heat capacities that are much greater than that of the overlying envelope. This condition holds for $\hhe$ envelopes when the atmospheric mass fraction $M_\mathrm{env} / M_p \lesssim 4$ per cent, and is true for secondary atmospheric species like $\htwoo$ when $M_\mathrm{env} / M_p \lesssim 30$ per cent \citep{2019MNRAS.485.4454B}.
In this regime, significant mass loss can occur when the radius of the RCB is comparable to the outer radius, $R_\rcb \sim R_\mathrm{out}$. This is because energy lost from the envelope as it escapes can be resupplied by the hot planetary surface, maintaining the envelope in an inflated state that promotes loss. The required post-impact temperature at the base of the atmosphere is

\begin{align}
\label{eq: mu loss temperature}
T'_{p,\mathrm{crit}} = T_\mathrm{eq} \left[ 1 + \frac{\gamma - 1}{2} \left( \frac{R'_B}{R'_p} - \frac{R'_B}{R'_\mathrm{out}} \right) \right] \text{,}
\end{align}
where unprimed and primed quantities indicate the pre- and post-impact values, respectively (e.g., $M'_p = M_p + M_\mathrm{imp}$).

Because the Bondi radius $R_B \propto \mu$, the required temperature for significant atmospheric loss given in Equation \eqref{eq: mu loss temperature} increases with the mean molecular weight of the envelope. For an Earth-like planet with a $\hhe$ envelope ($\mu = 2.3;~ \gamma = 7/5$), the critical temperature is ${\sim}5000~\mathrm{K}$, while for a water vapor atmosphere ($\mu = 18.0;~ \gamma = 4/3$) it is ${\sim}3 \times 10^4 ~\mathrm{K}$. As $T'_{p,\mathrm{crit}}$ increases, so does the required impactor mass to achieve it. Writing the impact velocity in terms of the mutual escape velocity yields an impact energy of $E_\mathrm{imp} \simeq (v_\mathrm{imp}/v_\mathrm{esc})^2 G M_\mathrm{imp} (M_\mathrm{imp} + M_p) / (R_\mathrm{imp} + R_p)$. Combining this expression with the above temperature yields an equation for the largest mean molecular weight atmosphere for which an impactor can drive thermal loss:
\begin{align}
\label{eq: impact mu}
\mu \simeq & \frac{\gamma}{\gamma - 1} \left[ \eta \left(\frac{v_\mathrm{imp}}{v_\mathrm{esc}} \right)^2 \frac{M_\mathrm{imp}}{M_\mathrm{imp} + M_p} \frac{R'_p}{R_\mathrm{imp} + R_p} \frac{k_B}{m_p c_{V,p}} \right.
\\
& \qquad + \left. \left( 1 - \frac{T_\mathrm{eq}}{T_p} \right) \frac{k_B T_p R'_p}{G m_p (M_\mathrm{imp} + M_p)} \right] 
\times \left( 1 - \frac{R'_p}{R'_\mathrm{out}} \right)^{-1} \text{.} \nonumber
\end{align}
For Earth-like planets at ${\sim}1~\mathrm{au}$, $R_p \ll R_\mathrm{out}$, so that the final term in parentheses is ${\sim}1$.

Collisions between equal-mass proto-planets provide an upper bound on the mean molecular weight of an atmosphere which can be lost due its thermal expansion following a giant impact, $\mu_\mathrm{max}$. In such an impact, the maximum mean molecular weight is
\begin{align}
\label{eq: max mu}
\mu_{\mathrm{max}} \simeq 0.6 \times \frac{\gamma}{\gamma - 1} \left[ \eta \frac{1}{2} \frac{k_B}{m_p c_{V,c}} 
+ \left( 1 - \frac{T_\mathrm{eq}}{T_p} \right) \frac{k_B T_p R_p}{G M_p m_p} \right]
\end{align}
where we have assumed $R'_p \ll R'_\mathrm{out}$ and that the planet and impactor have a mass-radius relationship defined by $R_p / R_\oplus = (M_p / M_\oplus)^n$ where $n \simeq 1/4$ \citep{2006Icar..181..545V}.
If the initial surface temperature $T_p \ll T'_{p,\mathrm{crit}}$, the bracketed expression is dominated by the first term, implying that the limit on the mass of atmospheric particles which can be lost is fixed by the heat capacity of the planet. This is because the planet's heat capacity is much greater than the atmosphere's, so it primarily determines the atmospheric base temperature, profile, and escape rate that results from a given impact energy. This limit on atmospheric mean molecular weight is therefore largely insensitive to changes in other planetary parameters.
Assuming a typical impact scenario with $\eta = 0.5$ and $v_\mathrm{imp} = \sqrt{2} v_\mathrm{esc}$, an initial base temperature of $T_p = 2000~\mathrm{K}$, and a core heat capacity $c_{V,p} = 7.5 \times 10^6~\mathrm{erg}~\mathrm{g}^{-1}~\mathrm{K}^{-1}$ \citep{2001PhRvB..64d5123A}, we find that the maximum molecular weight envelope that can be lost from planets in the mass range $M_p = 0.1\text{--}1 M_\oplus$ is less than both the molecular weight of water ($\mu_\mathrm{H_2O} \simeq 18$) and of carbon dioxide ($\mu_\mathrm{CO_2} \simeq 44$), the two main constituents expected in an early outgassed atmosphere (see Figure 3).
Specifically, for an impact yielding an Earth-mass planet at $1~\mathrm{au}$, $\mu_\mathrm{max} = 14$. For such a moderate initial base temperature, the result is relatively insensitive to heliocentric distance and planet mass.

\begin{figure}
	\centering
	\includegraphics[width=0.5\textwidth]{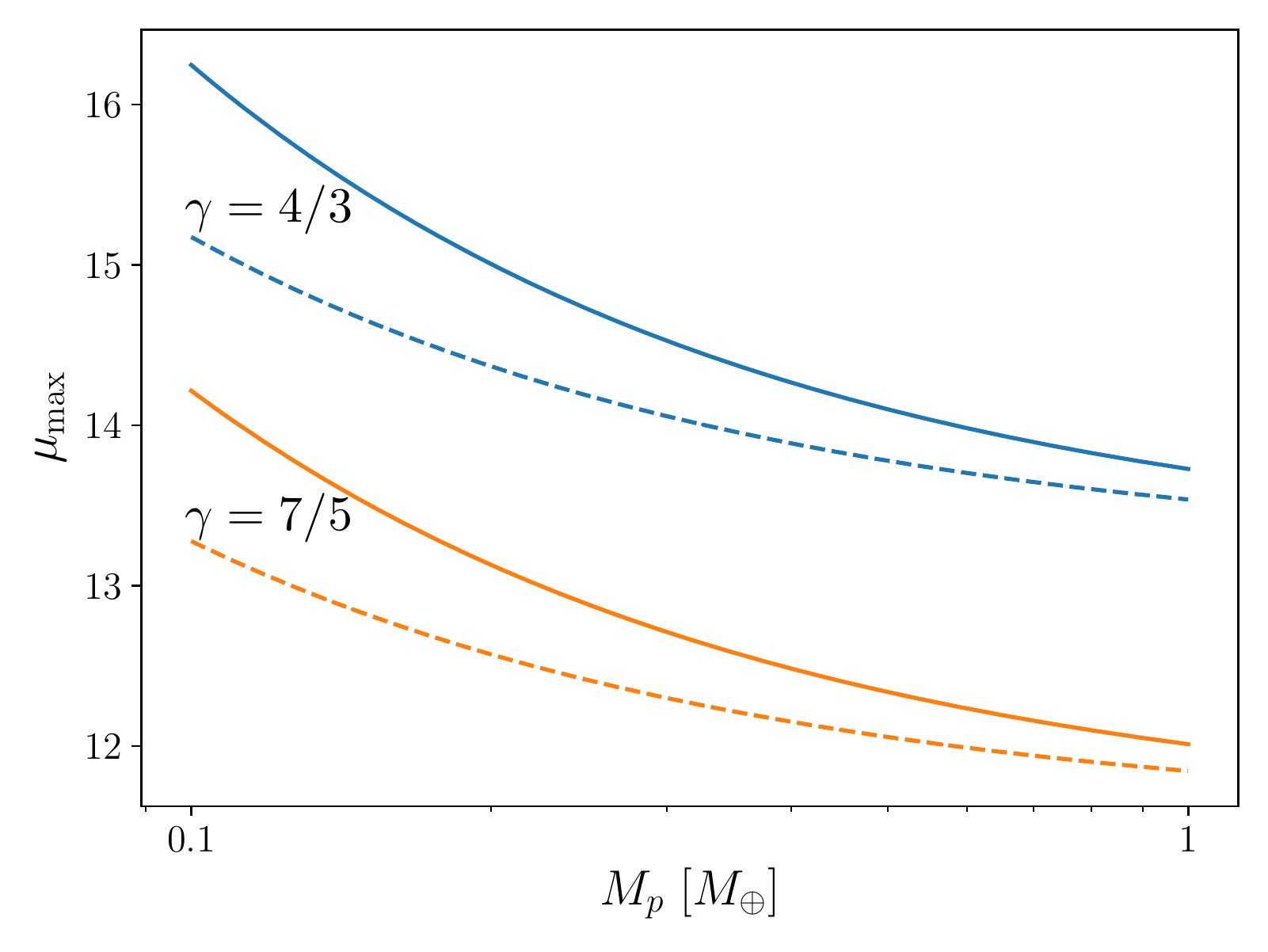}
	\caption{Maximum mean molecular weight envelope ($\mu_\mathrm{max}$) lost in an impact due to thermal expansion of the post-impact envelope as a function of initial target mass ($M_p$), calculated from Equation (1). The blue lines correspond to an adiabatic index for the envelope of $\gamma = 4/3$, while the orange lines correspond to $\gamma = 7/5$. Solid lines are for planets at $1~\mathrm{au}$ and dashed lines indicate $0.1~\mathrm{au}$. All planets have an initial base temperature, $T_p$, of $2000~\mathrm{K}$. The assumed impact velocity is  $v_\mathrm{imp} = \sqrt{2} v_\mathrm{esc}$, where $v_\mathrm{esc}$ is the mutual escape velocity of the impactor and target, the impactor mass is $M_\mathrm{imp} = M_p$, and the fraction of the impact energy that is converted to thermal energy is taken to be $\eta = 0.5$.}
	\label{fig: max mu}
\end{figure}

These calculations indicate that a pure secondary atmosphere is unlikely to be removed by thermal expansion after a giant impact. Secondary atmospheres may, however, still be removed by the hydrodynamic shock associated with the impact (see Section \ref{sec: terran numerical results}). 
Additionally, secondary components may be removed if they are mixed with lower molecular weight species. For example, a mixed $\hhe + \htwoo$ atmosphere with mass fractions of $f_\hhe = M_\hhe / M_p \sim 1$ per cent and $f_{\mathrm{H_2O}} \sim 0.1$ per cent would contain the equivalent of ${\sim}4$ Earth oceans of water, but have a mean molecular weight of $\mu \simeq 2.5$. If such an atmosphere is well-mixed to the outer radius, the above calculations would indicate it could be lost in a giant impact even though a pure steam atmosphere with the same water mass could not.

\subsection{Compositional differentiation and hydrodynamic drag}
\label{sec: hydro drag mass frac}
If an atmosphere is well-mixed at all altitudes, the relative abundances in the escaping gas are the same as those in bulk atmosphere. If this is not the case, then the preferential loss of one species over another leads to fractionation in the remaining atmosphere. This effect during the hydrodynamic escape of hydrogen has been previously studied and proposed as a possible explanation for the relative abundances of noble gases on the terrestrial planets \citep{1973JAtS...30.1481H, 1990Icar...84..502Z, 1991Icar...92....2P}. The extent of this effect depends on the efficiency of vertical mixing and the rate of atmospheric loss.

An atmosphere is vertically well-mixed when its rate of turbulent mixing is more rapid than molecular diffusion. The altitude where these processes become equally important is defined as the homopause.
Above the homopause molecular diffusion dominates and atmospheric species may segregate themselves according to their respective molecular weights. 
Therefore, the density of a heavier atmospheric component may be substantially depleted at high altitudes compared to its bulk composition in the atmosphere. We assume the atmosphere is well-mixed to the RCB, $R_\mathrm{homo} = R_\rcb$. This is both because vigorous convection below the RCB promotes turbulent mixing and because higher densities in this region slow the rate of molecular diffusion, which is ${\propto}n^{-1}$ where $n$ is the atmospheric number density. In a static atmosphere, we would therefore expect much lower concentrations of heavier species at altitudes much higher than the RCB.

In cases of rapid atmospheric escape, however, the escaping flow of light molecules can drag along heavier molecules, leading to their escape.
For a flux of a light species $F_i$, the flux of a heavier species can be written \citep{1990Icar...84..502Z}
\begin{align}
\label{eq: drag flux}
F_j = F_i x_j n_j / n_i \text{,}
\end{align}
where $n_j$ and $n_i$ are the number densities of species $j$ and $i$ and $x_j \in [0, 1]$ is the fractionation factor, which determines the efficiency with which heavier molecules are dragged by the flow. The fractionation factor is defined as
\begin{align}
\label{eq: fractionation factor}
x_j = 1 - \frac{g \left(\mu_j - \mu_i \right) m_p b_{i,j}}{F_i k_B T \left( 1 + n_j / n_i \right)} \text{,}
\end{align}
where $g$ is the gravitational acceleration and $b_{i,j}$ is the binary interaction cross-section. Radially varying quantities are evaluated at the base of the outflow region, which we take to be the RCB.

We include the effects of compositional differentiation and hydrodynamic drag in our calculations and track the behavior of individual species as the atmosphere evolves. The mass loss rate is still governed by Equation \eqref{eq: mdot definition}. In the energy-limited regime, the mass loss rate of each species is determined by simultaneously solving the above equation for the fractionation factor (Equation \eqref{eq: fractionation factor}) and for mass conservation ($\dot{M} = \dot{M}_i + \dot{M}_j$). In the supply-limited regime, the hydrogen flux, taken to be species $i$, in the fractionation factor is determined by $\dot{M}_i = 4 \pi R_\mathrm{out}^2 \rho_{i, \mathrm{out}}$. The density of hydrogen is calculated from
\begin{align}
\rho_i = \rho_{i, \rcb} \exp{\left[\frac{R_\rcb}{h_i} \left(\frac{R_\rcb}{r} - 1 \right) \right]} \text{,}
\end{align}
where, assuming $R_\mathrm{homo} = R_\rcb$ so that atmospheric species are diffusively separated, the scale height of the hydrogen is $h_i = k_B T_\mathrm{eq} R^2_\rcb / (G M_p \mu_i m_p)$. The total mass loss rate is then determined by combining the hydrogen flux with the flux of the heavier, dragged species determined by Equation \eqref{eq: drag flux}. We typically find that the high mass loss rates following an impact lead to fractionation factors $x_j \sim 1$, so that the relative flux rates of each species are proportional to their abundances in the total atmosphere. In other words, the mass loss proceeds similarly to the case where the atmosphere is assumed to be well-mixed to the outer radius ($R_\mathrm{homo} = R_\mathrm{out}$).

This can be understood as follows. Approximating $F_i = \dot{M}_i / (4\pi r^2 \mu_i m_p) \simeq \dot{M} / (4 \pi r^2 \mu m_p)$ in a hydrogen-dominated atmosphere, the fractionation factor can be rewritten as
\begin{align}
\label{eq: mod fractionation factor}
x_j & = 1 - \frac{G M_p}{\dot{M}} \frac{4 \pi \mu_i \left( \mu_j - \mu_i \right) m_p^2 b_{i,j}}{k_B T_\mathrm{eq} \left(1 + n_j / n_i \right)}
\\
\Rightarrow x_\htwoo & \sim 1 - \left( \frac{0.01}{f} \right) \left( \frac{\tau_\mathrm{loss}}{7~\mathrm{Gyr}} \right) \text{ and } 
\\
x_\cotwo & \sim 1 - \left( \frac{0.01}{f} \right) \left( \frac{\tau_\mathrm{loss}}{3~\mathrm{Gyr}} \right) \text{,}
\end{align}
where we have used $b_{\hhe, \htwoo} \simeq 2.7 \times 10^{17} T^{0.75} ~\mathrm{cm}^{-1}~\mathrm{s}^{-1}$ and $b_{\hhe, \cotwo} \simeq 2.3 \times 10^{17} T^{0.75} ~\mathrm{cm}^{-1}~\mathrm{s}^{-1}$ \citep{1986Icar...68..462Z}, and made the substitution $\dot{M} \sim M_\mathrm{env} / \tau_\mathrm{loss} \sim f M_p / \tau_\mathrm{loss}$. Following giant impacts, typical mass loss timescales for $\hhe$ envelopes with a mass fraction $f \sim 1$ per cent are ${\lesssim}100~\mathrm{Myr}$ \citep{2019MNRAS.485.4454B}.
Over these timescales, the fractionation factors for $\htwoo$ and $\cotwo$ are ${\sim}1$, with $\htwoo$ marginally more easily entrained in the hydrogen outflow. Because the source of the outflow is the well-mixed convective region, this implies that $F_\htwoo / n_\htwoo \simeq F_\hhe / n_\hhe$, implying that the total escape flux is similar to what is expected from a well-mixed envelope ($R_\mathrm{homo} = R_\mathrm{out}$).

\section{Numerical results}
\label{sec: terran numerical results}
To capture the full range of possible outcomes during the giant impact phase, we calculate the post-impact evolution of four example atmospheres, representing plausible atmospheric states during terrestrial planet formation, with the model described above. Specifically, we examine the loss of a pure $\hhe$ atmosphere, a steam atmosphere, and two mixed $\hhe + \htwoo$ atmospheres. In each case, we assume that the initial planet has $M_p = 1 M_\oplus$, $T_{c, 0} = 2000~\mathrm{K}$, and an orbital radius of $a = 1.0~\mathrm{au}$ around a sun-like star. We neglect any atmospheric mass contributed by the impactor and assume that the impact results in perfect accretion so that the new core mass is $M'_p = M_p + M_\mathrm{imp}$. 
Finally, we adopt $\eta = 0.5$ and $v_\mathrm{imp} = \sqrt{2} v_\mathrm{esc}$ (see Section \ref{sec: terran discussion}). 

\begin{figure}
\centering
	\includegraphics[width=0.5\textwidth]{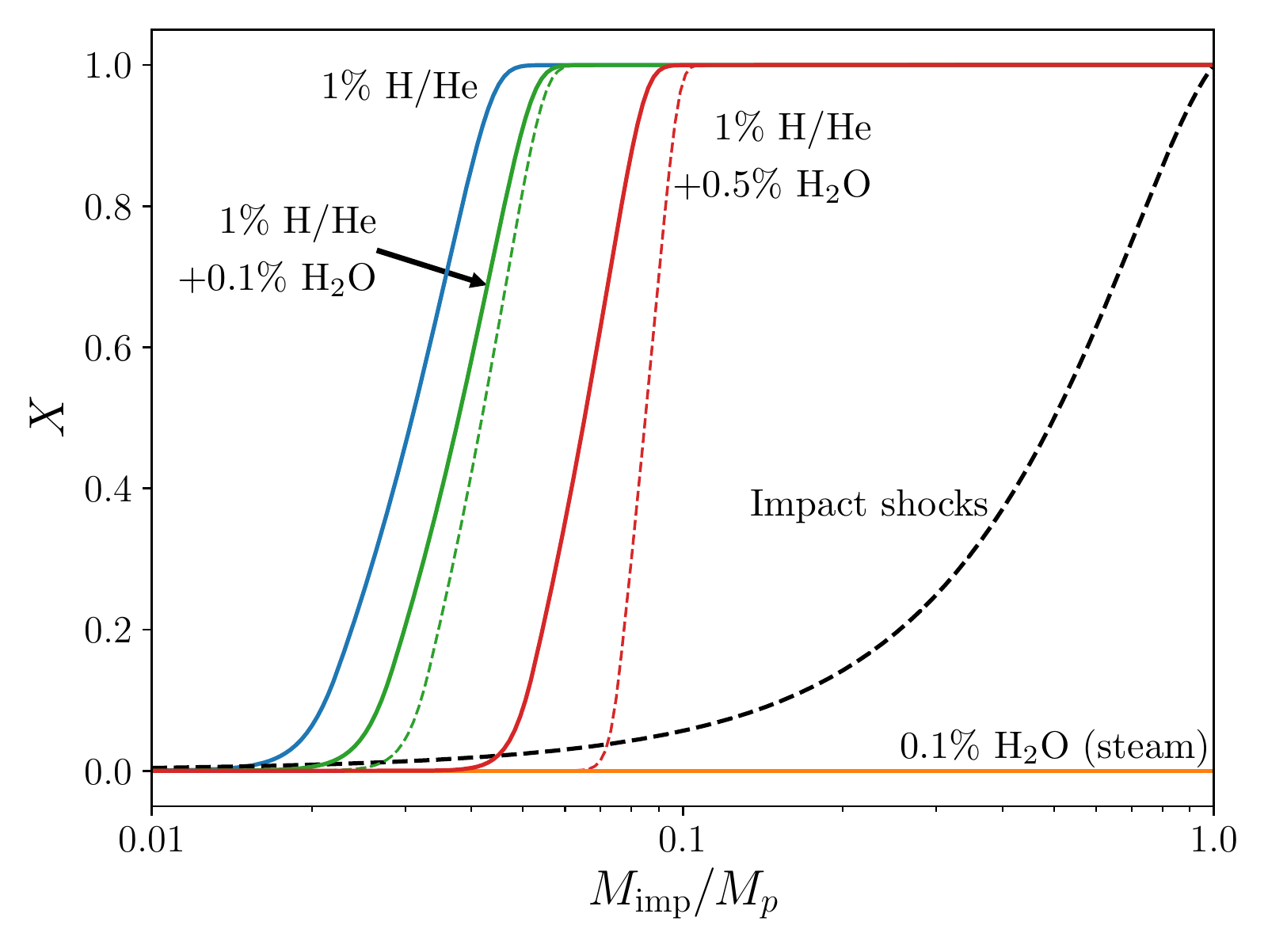}
	\caption{Mass fraction of envelope lost ($X$) as a function of impactor mass ($M_\mathrm{imp}/M_p$) for planets with $M_p = 1 M_\oplus$, $R_p = 1 R_\oplus$, $T_p = 2000~\mathrm{K}$, a = $1~\mathrm{au}$, and different envelope compositions and mass fractions (as indicated by their respective labels). The two green and red lines for $\hhe + \htwoo$ envelope compositions represent the two different mixing models: the bold solid line corresponds to hydrodynamic escape with mass fractionation, while the thin dashed line is for a well-mixed envelope. The black dashed line indicates the atmospheric envelope loss from the impact shock for $v_\mathrm{imp} \simeq v_\mathrm{esc}$ \citep[][Equation (33)]{2015Icar..247...81S}.}
	\label{fig: terran mass loss v imp. mass}
\end{figure}

For the primordial atmosphere case, we assume an initial $\hhe$ envelope with a mass fraction $f = M_\mathrm{env} / M_p = 1$ per cent. This approximately corresponds to the envelope mass that an Earth-mass planet could accrete if it grew to ${\gtrsim}50$ per cent of its current mass prior to the dissipation of the gas disk \citep{2014MNRAS.439.3225L, 2015ApJ...811...41L, 2015MNRAS.448.1751I}. We find, consistent with previous work \citep{2019MNRAS.485.4454B}, that the $\hhe$ envelope is readily lost during giant impacts for the impactor mass fractions $M_\mathrm{imp} / M_p \gtrsim 0.05$ (Figure \ref{fig: terran mass loss v imp. mass}), implying that a Mars-sized impactor hitting the proto-Earth would have ejected the entire H/He envelope.

The next case we consider is a secondary atmosphere entirely consisting of outgassed $\htwoo$ (orange line in Figure \ref{fig: terran mass loss v imp. mass}). Here we assume an atmospheric mass of $f = 0.1$ per cent, which is equivalent in mass to about four Earth oceans, a moderate estimate for the outgassed water content of the magma ocean \citep{2008E&PSL.271..181E, 2013Natur.497..607H}. In contrast to a primordial $\hhe$ atmosphere, the pure secondary steam atmosphere undergoes no appreciable atmospheric loss due to the heating and thermal expansion of the envelope following the impact in any of the scenarios considered here.

Most notable are the atmospheric mass-loss results for secondary atmospheres that are outgassed into a hydrogen-dominated envelope (green and red lines in Figure \ref{fig: terran mass loss v imp. mass}). To model this scenario, we consider two atmospheric compositions. First, we combine the primary and secondary atmospheres described above for a total mass fraction $f = 1.1$ per cent. For such a mixed atmosphere, we find complete atmospheric loss for impactor masses exceeding about $5$ per cent of the target mass (green line in Figure \ref{fig: terran mass loss v imp. mass}). Second, we increase the water vapor fraction of the atmosphere to the equivalent of 20 Earth oceans. Even then, a typical giant impact results in efficient loss of the atmosphere (red line in Figure \ref{fig: terran mass loss v imp. mass}). These results can be understood by examining the mean molecular weights of the mixed atmospheres. The mean molecular weights of the atmospheres with ${\sim}4$ and $20$ oceans' worth of water vapor are $\mu \simeq 2.5$ and $3.3$, respectively.
In both cases, this is significantly less than the maximum mean molecular weight atmosphere that can be lost given by Equation \eqref{eq: max mu}. As expected, the higher mean molecular weight of the mixed atmospheres does require slightly larger impactors to achieve the same loss fraction as pure $\hhe$ atmospheres (see Equation \eqref{eq: impact mu}). Additionally, we find that simulations using a well-mixed atmosphere model require slightly larger impactors to achieve atmospheric loss comparable to that in simulations using the fractionation model (dashed and solid green lines, respectively, in Figure \ref{fig: terran mass loss v imp. mass}). This is because the atmospheric hydrogen is more extended in the case where the radiative region is un-mixed ($h_\hhe > h$), lowering the impactor mass needed to initiate atmospheric loss. Because the loss of heavier species is driven by the hydrogen flux, heavier molecules are also lost more easily in the un-mixed case, despite their much smaller scale heights.

Finally, comparison with the estimated shock-induced losses from \citet{2015Icar..247...81S} indicates that atmospheric loss from thermal expansion dominates the primordial and mixed envelopes for most of the impactor mass range considered, with the two processes becoming comparable only when $M_\mathrm{imp} \sim M_p$. For the secondary atmosphere, however, shocks are the more important process at all impactor masses.

\section{Discussion and conclusions}
\label{sec: terran discussion}
\begin{figure*}
	\includegraphics[width=\textwidth]{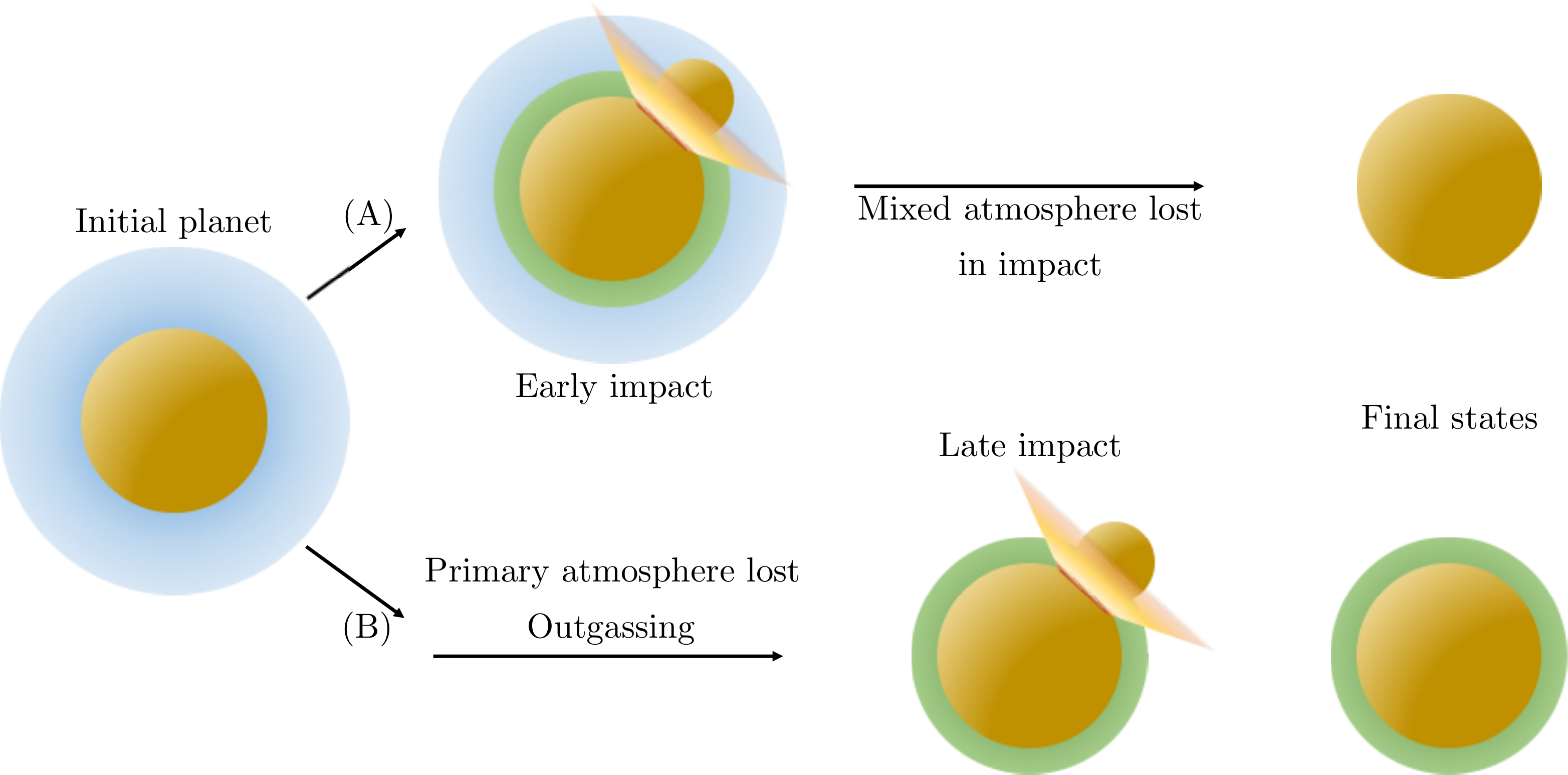}
	\caption{\small Example of divergent atmospheric outcomes following a collision at different times during the giant impact phase. Planets A and B (top and bottom rows, respectively) begin with identical primary atmospheres (shown in blue) and outgas identical secondary atmospheres (depicted in green). However, planet A loses its mixed primordial-secondary atmosphere during an impact and is left with a negligible atmosphere after the impact and with a significantly depleted volatile budget. Planet B suffers an impact after the primary atmosphere is lost by either photoevaporation or an earlier impact such that it has a pure secondary atmosphere at the time of the giant impact depicted in the figure. In this case, it retains most of its secondary atmosphere and hence most of its volatiles during subsequent giant impacts.}
	\label{fig: terran pathways}
\end{figure*}

The above case studies allow us to draw several broad conclusions about the potential role of giant impacts in the atmospheric evolution of terrestrial planets. Planets with light hydrogen-dominated envelopes can lose the bulk of their atmospheres through a planetary wind following a large impact ($M_\mathrm{imp} / M_p \gtrsim 0.05$). Thermal loss following an impact can also deplete the inventory of high mean molecular weight volatiles, provided they are sufficiently diluted by $\hhe$ in the bulk atmosphere.
These results are remarkable in that they provide an efficient way to lose several oceans' worth of water and other volatile species that are predicted to be present in the atmospheres of terrestrial planets as a result of outgassing. Thermal atmospheric loss triggered by giant impacts can therefore radically alter the volatile budget of terrestrial planets. Total atmospheric loss in these scenarios would likely be enhanced by impact shocks and high-energy stellar irradiation, so that the cumulative loss after an impact may be even greater \citep[e.g.,][]{2012ApJ...761...59L, 2015MNRAS.448.1751I, 2015ApJ...812..164L, 2016MNRAS.460.1300E}.

Atmospheric loss following a giant impact is highly dependent on the atmosphere's composition.
For impactors with $v_\mathrm{imp} \sim v_\mathrm{esc}$, we find that thermal escape of a purely outgassed secondary atmosphere following an impact is negligible, and that the total impact-driven loss is dominated by shocks. We do not model impact shocks in this work, but previous studies suggest that, for impactors with $v_\mathrm{imp} \sim v_\mathrm{esc}$, removal of the entire envelope requires $M_\mathrm{imp} / M_p \sim 1$ \citep{2015Icar..247...81S, 2015MNRAS.448.1751I}.
In the canonical moon-forming impact ($M_\mathrm{imp} / M_p \sim 0.1$), atmospheric losses of ${\sim}20$ per cent are expected due to shock, though this can be enhanced for some choices of the proto-Earth's initial conditions \citep{2003Icar..164..149G, 2005Natur.433..842G} and scenarios with larger impactor masses ($M_\mathrm{imp} / M_p \sim 1$) have been proposed \citep{2012Sci...338.1052C}.
Therefore, while some scenarios may produce significant erosion of secondary atmospheres, these atmospheres are generally difficult to dislodge in a single giant impact event.

In addition to atmospheric composition, the outcome of a giant impact depends on the details of the impact scenario, in particular the impactor size ($M_\mathrm{imp} / M_p$), impact velocity ($v_\mathrm{imp}$), and timing. Terrestrial planet formation likely encompasses a broad array of plausible scenarios and the details for any particular planet are stochastic, potentially producing dramatically different results. In simulations adopting either the current orbits of the giant planets or a more compact and circular configuration, impactor mass ratios range from $M_\mathrm{imp} / M_p \sim 0.1\text{--}0.3$, impact velocities are ${\sim}1\text{--}1.3 v_\mathrm{esc}$, and impact times are ${\sim}10\text{--}100~\mathrm{Myr}$ after the formation of the solar system \citep[e.g.,][]{2006Icar..184...39O}. In the ``Grand Tack'' scenario, characterized by inward then outward migration of Jupiter in the presence of the gas disk \citep{2011Natur.475..206W}, a wider range of impact scenarios occur: $M_\mathrm{imp}/M_p \sim 0.01\text{--}0.3$, $v_\mathrm{imp} \sim 1\text{--}2 v_\mathrm{esc}$, and impact times of ${\sim}0.1\text{--}100~\mathrm{Myr}$ \citep{2014Icar..239...74O}. In a model including both the dissipating disk and the growth of planetary embryos, \citet{2019Icar..329...88W} find a delayed onset for the dynamical instability with impacts beginning at ${\sim}10\text{--}20~\mathrm{Myr}$. For an estimated value of $\eta \simeq 0.5$ \citep{2020JGRE..12506042C} most of these impacts can be expected to entirely remove $\hhe$ envelopes. The only exceptions are the impactors at the low end of the mass range in the ``Grand Tack'' scenario. Therefore, while similar planets may experience a diversity of impact scenarios, it appears that the most plausible range of impactor masses and velocities will efficiently remove primordial $\hhe$ envelopes as well as envelopes that consist of a mix of primordial hydrogen and secondary outgassed species.

The variable timing of impacts, however, may result in significantly different evolutionary pathways for planets. This is because the giant impact phase occurs concurrently with multiple other evolutionary processes for terrestrial planet atmospheres. 
Specifically, a proto-planet with a large fraction of its volatiles mixed into a hydrogen-dominated atmosphere may lose its entire atmosphere in a giant impact, resulting in a substantially depleted volatile inventory. On the other hand, an identical proto-planet that lost its primordial atmosphere either by an early giant impact or through high-energy stellar radiation \citep{2015ApJ...815L..12J} prior to outgassing of a secondary atmosphere will be able to retain most of its secondary atmosphere and, hence, its volatile budget, for the duration of the giant impact phase (see Figure \ref{fig: terran pathways}). Early removal of a hydrogen envelope by an impact may allow a planet's magma ocean to cool and outgas earlier than it otherwise would. The components of this outgassed atmosphere might then face greater erosion from increased cumulative exposure to planetesimal impacts and high-energy stellar radiation \citep{2015Icar..247...81S}.

In the specific case of the solar system, evidence for the accretion of Mars on a timescale comparable to the gas disk lifetime \citep{2011Natur.473..489D} suggests the planetary embryos that formed the inner planets grew quickly enough to accrete a primordial atmosphere. The composition of noble gases in the Earth's mantle is consistent with the dissolution of gas from such an atmosphere into the early Earth's magma ocean \citep{1996Sci...273.1814H, 2012Natur.486..101M, 2019Natur.565...78W}. If this is the case, the atmospheric evolution of these planets may have been significantly affected by giant impacts. This scenario is consistent with xenon abundance measurements that suggest the proto-Earth once hosted roughly four times its presently observed volatile inventory \citep{2014RSPTA.37230260A} and therefore could offer an explanation for the depletion of Earth's and Mars's light noble gas budgets compared to Venus \citep{1997Icar..126..148P, 2013GeCoA.105..146H}.

Giant impacts are an expected component of terrestrial planet formation, and have the potential to significantly shape the volatile inventories of accreting terrestrial planets. We find that the outcome of an impact depends strongly on the atmospheric composition. Low mean molecular weight envelopes can be efficiently removed by a planetary wind following these events, and loss by this mechanism significantly exceeds that caused by the initial impact shock. Additionally, the rate of hydrogen escape following an impact is so rapid that large reservoirs of high mean molecular weight volatiles can be lost if they are sufficiently diluted in a hydrogen-dominated atmosphere. In contrast, purely secondary atmospheres (e.g., steam envelopes) are only moderately diminished during giant impacts, unless $M_\mathrm{imp} \sim M_p$. For these atmospheres, loss is due to the impact shock and thermal loss following the impact is negligible.
Because giant impacts occur stochastically during a time when nascent planetary atmospheres are being altered by other processes, including outgassing and photoevaporation, these impacts likely play a key role in determining the ultimate volatile inventories of terrestrial planets, leading to a large diversity in composition among otherwise similar worlds.

\section*{Acknowledgements}
This research has been supported by the National Aeronautics and Space Administration under grant No. $17~\rm{XRP}17\_~2-0055$ issued through the Exoplanet Research Program. This research made use of the software packages NumPy \citep{van2011numpy}, SciPy \citep{2020SciPy-NMeth}, Astropy \citep{2013A&A...558A..33A}, and matplotlib \citep{Hunter:2007}.


\bibliographystyle{mnras}
\bibliography{GiantImpact2}

\begin{thebibliography}{}
\makeatletter
\relax
\def\mn@urlcharsother{\let\do\@makeother \do\$\do\&\do\#\do\^\do\_\do\%\do\~}
\def\mn@doi{\begingroup\mn@urlcharsother \@ifnextchar [ {\mn@doi@}
  {\mn@doi@[]}}
\def\mn@doi@[#1]#2{\def\@tempa{#1}\ifx\@tempa\@empty \href
  {http://dx.doi.org/#2} {doi:#2}\else \href {http://dx.doi.org/#2} {#1}\fi
  \endgroup}
\def\mn@eprint#1#2{\mn@eprint@#1:#2::\@nil}
\def\mn@eprint@arXiv#1{\href {http://arxiv.org/abs/#1} {{\tt arXiv:#1}}}
\def\mn@eprint@dblp#1{\href {http://dblp.uni-trier.de/rec/bibtex/#1.xml}
  {dblp:#1}}
\def\mn@eprint@#1:#2:#3:#4\@nil{\def\@tempa {#1}\def\@tempb {#2}\def\@tempc
  {#3}\ifx \@tempc \@empty \let \@tempc \@tempb \let \@tempb \@tempa \fi \ifx
  \@tempb \@empty \def\@tempb {arXiv}\fi \@ifundefined
  {mn@eprint@\@tempb}{\@tempb:\@tempc}{\expandafter \expandafter \csname
  mn@eprint@\@tempb\endcsname \expandafter{\@tempc}}}

\bibitem[\protect\citeauthoryear{{Alf{\`e}}, {Price}  \& {Gillan}}{{Alf{\`e}}
  et~al.}{2001}]{2001PhRvB..64d5123A}
{Alf{\`e}} D.,  {Price} G.~D.,   {Gillan} M.~J.,  2001, \mn@doi [\prb]
  {10.1103/PhysRevB.64.045123}, \href
  {http://adsabs.harvard.edu/abs/2001PhRvB..64d5123A} {64, 045123}

\bibitem[\protect\citeauthoryear{{Asphaug}}{{Asphaug}}{2014}]{2014AREPS..42..551A}
{Asphaug} E.,  2014, \mn@doi [Annual Review of Earth and Planetary Sciences]
  {10.1146/annurev-earth-050212-124057}, \href
  {https://ui.adsabs.harvard.edu/abs/2014AREPS..42..551A} {42, 551}

\bibitem[\protect\citeauthoryear{{Astropy Collaboration} et~al.,}{{Astropy
  Collaboration} et~al.}{2013}]{2013A&A...558A..33A}
{Astropy Collaboration} et~al., 2013, \mn@doi [\aap]
  {10.1051/0004-6361/201322068}, \href
  {http://adsabs.harvard.edu/abs/2013A%26A...558A..33A} {558, A33}

\bibitem[\protect\citeauthoryear{{Avice} \& {Marty}}{{Avice} \&
  {Marty}}{2014}]{2014RSPTA.37230260A}
{Avice} G.,  {Marty} B.,  2014, \mn@doi [Philosophical Transactions of the
  Royal Society of London Series A] {10.1098/rsta.2013.0260}, \href
  {https://ui.adsabs.harvard.edu/abs/2014RSPTA.37230260A} {372, 20130260}

\bibitem[\protect\citeauthoryear{{Barboni}, {Boehnke}, {Keller}, {Kohl},
  {Schoene}, {Young}  \& {McKeegan}}{{Barboni}
  et~al.}{2017}]{2017SciA....3E2365B}
{Barboni} M.,  {Boehnke} P.,  {Keller} B.,  {Kohl} I.~E.,  {Schoene} B.,
  {Young} E.~D.,   {McKeegan} K.~D.,  2017, \mn@doi [Science Advances]
  {10.1126/sciadv.1602365}, \href
  {https://ui.adsabs.harvard.edu/abs/2017SciA....3E2365B} {3, e1602365}

\bibitem[\protect\citeauthoryear{{Biersteker} \& {Schlichting}}{{Biersteker} \&
  {Schlichting}}{2019}]{2019MNRAS.485.4454B}
{Biersteker} J.~B.,  {Schlichting} H.~E.,  2019, \mn@doi [\mnras]
  {10.1093/mnras/stz738}, \href
  {https://ui.adsabs.harvard.edu/abs/2019MNRAS.485.4454B} {485, 4454}

\bibitem[\protect\citeauthoryear{{Canup}}{{Canup}}{2012}]{2012Sci...338.1052C}
{Canup} R.~M.,  2012, \mn@doi [Science] {10.1126/science.1226073}, \href
  {https://ui.adsabs.harvard.edu/abs/2012Sci...338.1052C} {338, 1052}

\bibitem[\protect\citeauthoryear{{Canup} \& {Asphaug}}{{Canup} \&
  {Asphaug}}{2001}]{2001Natur.412..708C}
{Canup} R.~M.,  {Asphaug} E.,  2001, \nat, \href
  {https://ui.adsabs.harvard.edu/abs/2001Natur.412..708C} {412, 708}

\bibitem[\protect\citeauthoryear{{Carter}, {Lock}  \& {Stewart}}{{Carter}
  et~al.}{2020}]{2020JGRE..12506042C}
{Carter} P.~J.,  {Lock} S.~J.,   {Stewart} S.~T.,  2020, \mn@doi [Journal of
  Geophysical Research (Planets)] {10.1029/2019JE006042}, \href
  {https://ui.adsabs.harvard.edu/abs/2020JGRE..12506042C} {125, e06042}

\bibitem[\protect\citeauthoryear{{Chambers} \& {Wetherill}}{{Chambers} \&
  {Wetherill}}{1998}]{1998Icar..136..304C}
{Chambers} J.~E.,  {Wetherill} G.~W.,  1998, \mn@doi [\icarus]
  {10.1006/icar.1998.6007}, \href
  {https://ui.adsabs.harvard.edu/abs/1998Icar..136..304C} {136, 304}

\bibitem[\protect\citeauthoryear{{Dauphas} \& {Pourmand}}{{Dauphas} \&
  {Pourmand}}{2011}]{2011Natur.473..489D}
{Dauphas} N.,  {Pourmand} A.,  2011, \mn@doi [\nat] {10.1038/nature10077},
  \href {https://ui.adsabs.harvard.edu/abs/2011Natur.473..489D} {473, 489}

\bibitem[\protect\citeauthoryear{{Elkins-Tanton}}{{Elkins-Tanton}}{2008}]{2008E&PSL.271..181E}
{Elkins-Tanton} L.~T.,  2008, \mn@doi [Earth and Planetary Science Letters]
  {10.1016/j.epsl.2008.03.062}, \href
  {https://ui.adsabs.harvard.edu/abs/2008E&PSL.271..181E} {271, 181}

\bibitem[\protect\citeauthoryear{{Elkins-Tanton}}{{Elkins-Tanton}}{2011}]{2011Ap&SS.332..359E}
{Elkins-Tanton} L.~T.,  2011, \mn@doi [\apss] {10.1007/s10509-010-0535-3},
  \href {https://ui.adsabs.harvard.edu/abs/2011Ap&SS.332..359E} {332, 359}

\bibitem[\protect\citeauthoryear{{Elkins-Tanton}}{{Elkins-Tanton}}{2012}]{2012AREPS..40..113E}
{Elkins-Tanton} L.~T.,  2012, \mn@doi [Annual Review of Earth and Planetary
  Sciences] {10.1146/annurev-earth-042711-105503}, \href
  {https://ui.adsabs.harvard.edu/abs/2012AREPS..40..113E} {40, 113}

\bibitem[\protect\citeauthoryear{{Erkaev}, {Lammer}, {Odert}, {Kislyakova},
  {Johnstone}, {G{\"u}del}  \& {Khodachenko}}{{Erkaev}
  et~al.}{2016}]{2016MNRAS.460.1300E}
{Erkaev} N.~V.,  {Lammer} H.,  {Odert} P.,  {Kislyakova} K.~G.,  {Johnstone}
  C.~P.,  {G{\"u}del} M.,   {Khodachenko} M.~L.,  2016, \mn@doi [\mnras]
  {10.1093/mnras/stw935}, \href
  {https://ui.adsabs.harvard.edu/abs/2016MNRAS.460.1300E} {460, 1300}

\bibitem[\protect\citeauthoryear{{Freedman}, {Marley}  \& {Lodders}}{{Freedman}
  et~al.}{2008}]{2008ApJS..174..504F}
{Freedman} R.~S.,  {Marley} M.~S.,   {Lodders} K.,  2008, \mn@doi [\apjs]
  {10.1086/521793}, \href {http://adsabs.harvard.edu/abs/2008ApJS..174..504F}
  {174, 504}

\bibitem[\protect\citeauthoryear{{Fressin} et~al.,}{{Fressin}
  et~al.}{2013}]{2013ApJ...766...81F}
{Fressin} F.,  et~al., 2013, \mn@doi [\apj] {10.1088/0004-637X/766/2/81}, \href
  {https://ui.adsabs.harvard.edu/abs/2013ApJ...766...81F} {766, 81}

\bibitem[\protect\citeauthoryear{{Genda} \& {Abe}}{{Genda} \&
  {Abe}}{2003}]{2003Icar..164..149G}
{Genda} H.,  {Abe} Y.,  2003, \mn@doi [\icarus]
  {10.1016/S0019-1035(03)00101-5}, \href
  {https://ui.adsabs.harvard.edu/abs/2003Icar..164..149G} {164, 149}

\bibitem[\protect\citeauthoryear{{Genda} \& {Abe}}{{Genda} \&
  {Abe}}{2005}]{2005Natur.433..842G}
{Genda} H.,  {Abe} Y.,  2005, \mn@doi [\nat] {10.1038/nature03360}, \href
  {https://ui.adsabs.harvard.edu/abs/2005Natur.433..842G} {433, 842}

\bibitem[\protect\citeauthoryear{{Ginzburg}, {Schlichting}  \&
  {Sari}}{{Ginzburg} et~al.}{2016}]{2016ApJ...825...29G}
{Ginzburg} S.,  {Schlichting} H.~E.,   {Sari} R.,  2016, \mn@doi [\apj]
  {10.3847/0004-637X/825/1/29}, \href
  {http://adsabs.harvard.edu/abs/2016ApJ...825...29G} {825, 29}

\bibitem[\protect\citeauthoryear{{Guillot}, {Chabrier}, {Gautier}  \&
  {Morel}}{{Guillot} et~al.}{1995}]{1995ApJ...450..463G}
{Guillot} T.,  {Chabrier} G.,  {Gautier} D.,   {Morel} P.,  1995, \mn@doi
  [\apj] {10.1086/176156}, \href
  {https://ui.adsabs.harvard.edu/abs/1995ApJ...450..463G} {450, 463}

\bibitem[\protect\citeauthoryear{{Halliday}}{{Halliday}}{2013}]{2013GeCoA.105..146H}
{Halliday} A.~N.,  2013, \mn@doi [\gca] {10.1016/j.gca.2012.11.015}, \href
  {https://ui.adsabs.harvard.edu/abs/2013GeCoA.105..146H} {105, 146}

\bibitem[\protect\citeauthoryear{{Hamano}, {Abe}  \& {Genda}}{{Hamano}
  et~al.}{2013}]{2013Natur.497..607H}
{Hamano} K.,  {Abe} Y.,   {Genda} H.,  2013, \mn@doi [\nat]
  {10.1038/nature12163}, \href
  {https://ui.adsabs.harvard.edu/abs/2013Natur.497..607H} {497, 607}

\bibitem[\protect\citeauthoryear{{Harper} \& {Jacobsen}}{{Harper} \&
  {Jacobsen}}{1996}]{1996Sci...273.1814H}
{Harper} Charles~L. J.,  {Jacobsen} S.~B.,  1996, \mn@doi [Science]
  {10.1126/science.273.5283.1814}, \href
  {https://ui.adsabs.harvard.edu/abs/1996Sci...273.1814H} {273, 1814}

\bibitem[\protect\citeauthoryear{{Hayashi}, {Nakazawa}  \& {Mizuno}}{{Hayashi}
  et~al.}{1979}]{1979E&PSL..43...22H}
{Hayashi} C.,  {Nakazawa} K.,   {Mizuno} H.,  1979, \mn@doi [Earth and
  Planetary Science Letters] {10.1016/0012-821X(79)90152-3}, \href
  {https://ui.adsabs.harvard.edu/abs/1979E&PSL..43...22H} {43, 22}

\bibitem[\protect\citeauthoryear{{Hunten}}{{Hunten}}{1973}]{1973JAtS...30.1481H}
{Hunten} D.~M.,  1973, \mn@doi [Journal of Atmospheric Sciences]
  {10.1175/1520-0469(1973)030<1481:TEOLGF>2.0.CO;2}, \href
  {https://ui.adsabs.harvard.edu/abs/1973JAtS...30.1481H} {30, 1481}

\bibitem[\protect\citeauthoryear{Hunter}{Hunter}{2007}]{Hunter:2007}
Hunter J.~D.,  2007, Computing In Science \& Engineering, 9, 90

\bibitem[\protect\citeauthoryear{{Inamdar} \& {Schlichting}}{{Inamdar} \&
  {Schlichting}}{2015}]{2015MNRAS.448.1751I}
{Inamdar} N.~K.,  {Schlichting} H.~E.,  2015, \mn@doi [\mnras]
  {10.1093/mnras/stv030}, \href
  {http://adsabs.harvard.edu/abs/2015MNRAS.448.1751I} {448, 1751}

\bibitem[\protect\citeauthoryear{{Izidoro}, {Raymond}, {Morbidelli}  \&
  {Winter}}{{Izidoro} et~al.}{2015}]{2015MNRAS.453.3619I}
{Izidoro} A.,  {Raymond} S.~N.,  {Morbidelli} A.~r.,   {Winter} O.~C.,  2015,
  \mn@doi [\mnras] {10.1093/mnras/stv1835}, \href
  {https://ui.adsabs.harvard.edu/abs/2015MNRAS.453.3619I} {453, 3619}

\bibitem[\protect\citeauthoryear{{Johansen} \& {Lambrechts}}{{Johansen} \&
  {Lambrechts}}{2017}]{2017AREPS..45..359J}
{Johansen} A.,  {Lambrechts} M.,  2017, \mn@doi [Annual Review of Earth and
  Planetary Sciences] {10.1146/annurev-earth-063016-020226}, \href
  {https://ui.adsabs.harvard.edu/abs/2017AREPS..45..359J} {45, 359}

\bibitem[\protect\citeauthoryear{{Johnstone} et~al.,}{{Johnstone}
  et~al.}{2015}]{2015ApJ...815L..12J}
{Johnstone} C.~P.,  et~al., 2015, \mn@doi [\apj] {10.1088/2041-8205/815/1/L12},
  \href {https://ui.adsabs.harvard.edu/abs/2015ApJ...815L..12J} {815, L12}

\bibitem[\protect\citeauthoryear{{Kite}, {Fegley}, {Schaefer}  \&
  {Ford}}{{Kite} et~al.}{2020}]{2020ApJ...891..111K}
{Kite} E.~S.,  {Fegley} Bruce J.,  {Schaefer} L.,   {Ford} E.~B.,  2020,
  \mn@doi [\apj] {10.3847/1538-4357/ab6ffb}, \href
  {https://ui.adsabs.harvard.edu/abs/2020ApJ...891..111K} {891, 111}

\bibitem[\protect\citeauthoryear{{Lammer} et~al.,}{{Lammer}
  et~al.}{2014}]{2014MNRAS.439.3225L}
{Lammer} H.,  et~al., 2014, \mn@doi [\mnras] {10.1093/mnras/stu085}, \href
  {https://ui.adsabs.harvard.edu/abs/2014MNRAS.439.3225L} {439, 3225}

\bibitem[\protect\citeauthoryear{{Lammer} et~al.,}{{Lammer}
  et~al.}{2018}]{2018A&ARv..26....2L}
{Lammer} H.,  et~al., 2018, \mn@doi [\aapr] {10.1007/s00159-018-0108-y}, \href
  {https://ui.adsabs.harvard.edu/abs/2018A&ARv..26....2L} {26, 2}

\bibitem[\protect\citeauthoryear{{Lee} \& {Chiang}}{{Lee} \&
  {Chiang}}{2015}]{2015ApJ...811...41L}
{Lee} E.~J.,  {Chiang} E.,  2015, \mn@doi [\apj] {10.1088/0004-637X/811/1/41},
  \href {http://adsabs.harvard.edu/abs/2015ApJ...811...41L} {811, 41}

\bibitem[\protect\citeauthoryear{{Liu}, {Hori}, {Lin}  \& {Asphaug}}{{Liu}
  et~al.}{2015}]{2015ApJ...812..164L}
{Liu} S.-F.,  {Hori} Y.,  {Lin} D.~N.~C.,   {Asphaug} E.,  2015, \mn@doi [\apj]
  {10.1088/0004-637X/812/2/164}, \href
  {http://adsabs.harvard.edu/abs/2015ApJ...812..164L} {812, 164}

\bibitem[\protect\citeauthoryear{{Lopez}, {Fortney}  \& {Miller}}{{Lopez}
  et~al.}{2012}]{2012ApJ...761...59L}
{Lopez} E.~D.,  {Fortney} J.~J.,   {Miller} N.,  2012, \mn@doi [\apj]
  {10.1088/0004-637X/761/1/59}, \href
  {http://adsabs.harvard.edu/abs/2012ApJ...761...59L} {761, 59}

\bibitem[\protect\citeauthoryear{{Morbidelli}, {Lunine}, {O'Brien}, {Raymond}
  \& {Walsh}}{{Morbidelli} et~al.}{2012}]{2012AREPS..40..251M}
{Morbidelli} A.,  {Lunine} J.~I.,  {O'Brien} D.~P.,  {Raymond} S.~N.,   {Walsh}
  K.~J.,  2012, \mn@doi [Annual Review of Earth and Planetary Sciences]
  {10.1146/annurev-earth-042711-105319}, \href
  {https://ui.adsabs.harvard.edu/abs/2012AREPS..40..251M} {40, 251}

\bibitem[\protect\citeauthoryear{{Mukhopadhyay}}{{Mukhopadhyay}}{2012}]{2012Natur.486..101M}
{Mukhopadhyay} S.,  2012, \mn@doi [\nat] {10.1038/nature11141}, \href
  {https://ui.adsabs.harvard.edu/abs/2012Natur.486..101M} {486, 101}

\bibitem[\protect\citeauthoryear{{O'Brien}, {Morbidelli}  \&
  {Levison}}{{O'Brien} et~al.}{2006}]{2006Icar..184...39O}
{O'Brien} D.~P.,  {Morbidelli} A.,   {Levison} H.~F.,  2006, \mn@doi [\icarus]
  {10.1016/j.icarus.2006.04.005}, \href
  {http://adsabs.harvard.edu/abs/2006Icar..184...39O} {184, 39}

\bibitem[\protect\citeauthoryear{{O'Brien}, {Walsh}, {Morbidelli}, {Raymond}
  \& {Mandell}}{{O'Brien} et~al.}{2014}]{2014Icar..239...74O}
{O'Brien} D.~P.,  {Walsh} K.~J.,  {Morbidelli} A.,  {Raymond} S.~N.,
  {Mandell} A.~M.,  2014, \mn@doi [\icarus] {10.1016/j.icarus.2014.05.009},
  \href {https://ui.adsabs.harvard.edu/abs/2014Icar..239...74O} {239, 74}

\bibitem[\protect\citeauthoryear{{Odert} et~al.,}{{Odert}
  et~al.}{2018}]{2018Icar..307..327O}
{Odert} P.,  et~al., 2018, \mn@doi [\icarus] {10.1016/j.icarus.2017.10.031},
  \href {https://ui.adsabs.harvard.edu/abs/2018Icar..307..327O} {307, 327}

\bibitem[\protect\citeauthoryear{{Pepin}}{{Pepin}}{1991}]{1991Icar...92....2P}
{Pepin} R.~O.,  1991, \mn@doi [\icarus] {10.1016/0019-1035(91)90036-S}, \href
  {https://ui.adsabs.harvard.edu/abs/1991Icar...92....2P} {92, 2}

\bibitem[\protect\citeauthoryear{{Pepin}}{{Pepin}}{1997}]{1997Icar..126..148P}
{Pepin} R.~O.,  1997, \mn@doi [\icarus] {10.1006/icar.1996.5639}, \href
  {https://ui.adsabs.harvard.edu/abs/1997Icar..126..148P} {126, 148}

\bibitem[\protect\citeauthoryear{{Piso} \& {Youdin}}{{Piso} \&
  {Youdin}}{2014}]{2014ApJ...786...21P}
{Piso} A.-M.~A.,  {Youdin} A.~N.,  2014, \mn@doi [\apj]
  {10.1088/0004-637X/786/1/21}, \href
  {https://ui.adsabs.harvard.edu/abs/2014ApJ...786...21P} {786, 21}

\bibitem[\protect\citeauthoryear{{Rafikov}}{{Rafikov}}{2006}]{2006ApJ...648..666R}
{Rafikov} R.~R.,  2006, \mn@doi [\apj] {10.1086/505695}, \href
  {https://ui.adsabs.harvard.edu/abs/2006ApJ...648..666R} {648, 666}

\bibitem[\protect\citeauthoryear{{Schlichting}, {Sari}  \&
  {Yalinewich}}{{Schlichting} et~al.}{2015}]{2015Icar..247...81S}
{Schlichting} H.~E.,  {Sari} R.,   {Yalinewich} A.,  2015, \mn@doi [\icarus]
  {10.1016/j.icarus.2014.09.053}, \href
  {http://adsabs.harvard.edu/abs/2015Icar..247...81S} {247, 81}

\bibitem[\protect\citeauthoryear{{Valencia}, {O'Connell}  \&
  {Sasselov}}{{Valencia} et~al.}{2006}]{2006Icar..181..545V}
{Valencia} D.,  {O'Connell} R.~J.,   {Sasselov} D.,  2006, \mn@doi [\icarus]
  {10.1016/j.icarus.2005.11.021}, \href
  {https://ui.adsabs.harvard.edu/#abs/2006Icar..181..545V} {181, 545}

\bibitem[\protect\citeauthoryear{{Valencia}, {Guillot}, {Parmentier}  \&
  {Freedman}}{{Valencia} et~al.}{2013}]{2013ApJ...775...10V}
{Valencia} D.,  {Guillot} T.,  {Parmentier} V.,   {Freedman} R.~S.,  2013,
  \mn@doi [\apj] {10.1088/0004-637X/775/1/10}, \href
  {https://ui.adsabs.harvard.edu/abs/2013ApJ...775...10V} {775, 10}

\bibitem[\protect\citeauthoryear{Van Der~Walt, Colbert  \& Varoquaux}{Van
  Der~Walt et~al.}{2011}]{van2011numpy}
Van Der~Walt S.,  Colbert S.~C.,   Varoquaux G.,  2011, Computing in Science \&
  Engineering, 13, 22

\bibitem[\protect\citeauthoryear{{Virtanen} et~al.,}{{Virtanen}
  et~al.}{2020}]{2020SciPy-NMeth}
{Virtanen} P.,  et~al., 2020, \mn@doi [Nature Methods]
  {https://doi.org/10.1038/s41592-019-0686-2}, \href {https://rdcu.be/b08Wh}
  {17, 261}

\bibitem[\protect\citeauthoryear{{Walsh} \& {Levison}}{{Walsh} \&
  {Levison}}{2019}]{2019Icar..329...88W}
{Walsh} K.~J.,  {Levison} H.~F.,  2019, \mn@doi [\icarus]
  {10.1016/j.icarus.2019.03.031}, \href
  {https://ui.adsabs.harvard.edu/abs/2019Icar..329...88W} {329, 88}

\bibitem[\protect\citeauthoryear{{Walsh}, {Morbidelli}, {Raymond}, {O'Brien}
  \& {Mandell}}{{Walsh} et~al.}{2011}]{2011Natur.475..206W}
{Walsh} K.~J.,  {Morbidelli} A.,  {Raymond} S.~N.,  {O'Brien} D.~P.,
  {Mandell} A.~M.,  2011, \mn@doi [\nat] {10.1038/nature10201}, \href
  {https://ui.adsabs.harvard.edu/abs/2011Natur.475..206W} {475, 206}

\bibitem[\protect\citeauthoryear{{Williams} \& {Mukhopadhyay}}{{Williams} \&
  {Mukhopadhyay}}{2019}]{2019Natur.565...78W}
{Williams} C.~D.,  {Mukhopadhyay} S.,  2019, \mn@doi [\nat]
  {10.1038/s41586-018-0771-1}, \href
  {https://ui.adsabs.harvard.edu/abs/2019Natur.565...78W} {565, 78}

\bibitem[\protect\citeauthoryear{{Zahnle} \& {Kasting}}{{Zahnle} \&
  {Kasting}}{1986}]{1986Icar...68..462Z}
{Zahnle} K.~J.,  {Kasting} J.~F.,  1986, \mn@doi [\icarus]
  {10.1016/0019-1035(86)90051-5}, \href
  {https://ui.adsabs.harvard.edu/abs/1986Icar...68..462Z} {68, 462}

\bibitem[\protect\citeauthoryear{{Zahnle}, {Kasting}  \& {Pollack}}{{Zahnle}
  et~al.}{1990}]{1990Icar...84..502Z}
{Zahnle} K.,  {Kasting} J.~F.,   {Pollack} J.~B.,  1990, \mn@doi [\icarus]
  {10.1016/0019-1035(90)90050-J}, \href
  {https://ui.adsabs.harvard.edu/abs/1990Icar...84..502Z} {84, 502}

\makeatother
\end{thebibliography}
\bsp
\label{lastpage}
\end{document}